# Scenarios for ILC in 2010

François Richard

*LAL, Univ Paris-Sud, IN2P3/CNRS, Orsay, France*

**Abstract**

Assuming that first significant results from LHC become available, this presentation assumes 4 different scenarios and discuss the implications for ILC

## Introduction

There is a common view concerning the relevance of the **Terascale** energy for providing a decisive insight on fundamental mechanisms governing our universe.

From what has been learned at present and past colliders (LEP/SLC and Tevatron) one expects that there should be a least a light Higgs detectable at LHC. This discovery will be a first and essential step to confirm our views on the origin of mass but it will take much more to provide a full explanation of the origin of the Higgs mechanism.

One could establish the origin of the Higgs by directly discovering new particles at LHC or by indirectly observing significant deviations in the various very precise observations allowed by the clean environment provided by ILC. In particular by measuring very precisely the decay modes of the Higgs bosons at ILC it should be possible to establish its true nature and the underlying mechanisms, SUSY or extra-dimensions, at work. A fascinating possibility, even challenging at ILC, will be to observe matter-anti matter asymmetry in the Higgs decays which would open an entirely new domain.

After LHC first results, those from Tevatron and non-accelerator various searches, we can expect, at the beginning of the next decade, the following scenarios:

- A  No signal with ~30 $fb^{-1}$ analyzed at LHC
- B  A Higgs found with a mass compatible with SM
- C  A Higgs found with a mass incompatible with SM and MSSM
- D  A Higgs has been found with non SM signals

In the following we will ask ourselves how can ILC at 500 GeV contribute to scenarios A,B,C,D?

## ILC in a nutshell

Before embarking in the discussion, let me summarize what is meant by ILC.

ILC will have a first phase at 500 GeV with polarized electrons, hopefully at the end of the next decade, and collect 500 $fb^{-1}$ in 4 years which is fully adequate to cover the Higgs SM and MSSM scenarios.

ILC can also measure top quark properties (electroweak couplings and mass) at the ‰ level. LHC will also copiously produce top quarks but dominantly through QCD processes.

Statistical accuracies reached at ILC will be unprecedented and it is therefore essential to work out the experimental strategy accordingly, avoiding from the start the usual limiting factors. We need a quasi perfect detector with high tracking accuracy, $dp/p^2 \sim 5 \cdot 10^{-5}$ and excellent jet reconstruction, $dE_j/E_j = 0.3 - 0.4/\sqrt{E_j}$ up to ~200 GeV jet energies. We need also these properties to be achieved down to ~100 mrad without any dead zones. b/c/t tagging capabilities will be improved with respect to SLC (and of course be far superior to LEP).

There will be no hardware trigger avoiding the difficult a priori choices which may bias unexpected physics.

4 concepts of detectors are under study and are thoroughly described in reference [1].

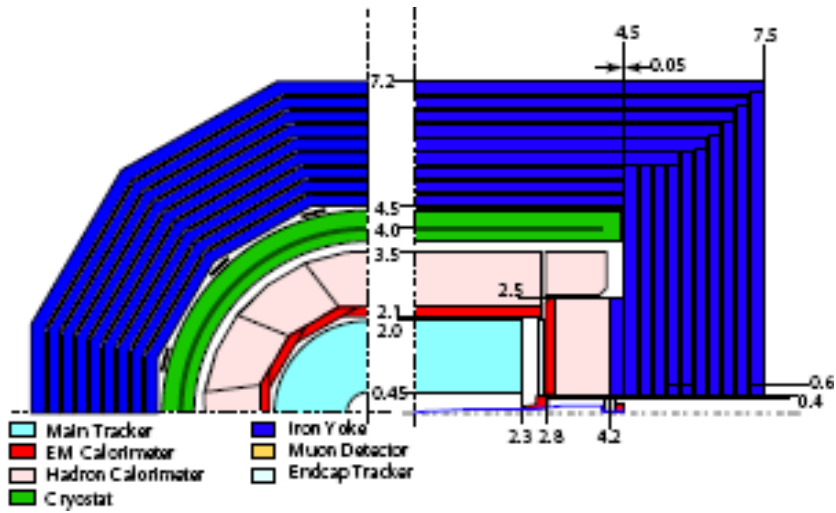

Figure 1: layout of one the concept detectors, the so called very large detector GLD.

Figure 1 shows one of them. The main feature used to achieve optimal jet reconstruction is to embed all calorimeters inside the magnetic coil therefore avoiding passive material transitions which usually spoil the reconstruction efficiency. Using integrated read-out micro-electronics it is also possible to reach an improved granularity for calorimetry. For tracking one benefits from improved measuring devices for the TPC and Si trackers.

Being able:
- to extract the differential luminosity $d\mathcal{L}/dE$ at better than 1‰
- to measure the beam energy E to better than 1‰ (for scans)
- to veto energetic e/$\gamma$ above ~5 mrad

are also important features needed in various analyses. An intense effort is underway to achieve these goals under what is called the MDI (Machine Detector Interface) framework which implies representatives of the detector concepts and machine experts.

## Scenario A

Recall that in this scenario there is no signal observed at LHC with ~30 fb$^{-1}$. We know that, within the SM, LHC should have observed a Higgs signal. Higgs particles may elude LHC searches in non minimal scenarios where SM cross-sections are reduced by a factor 3-5. ILC then provides the best possible detection for Higgs particles which can accommodate any non minimal scenario (NMSSM, CPV etc…) with reduced ZZH couplings. Recall also that LEP2 has not excluded such scenarios for mH<100 GeV. On the other hand one cannot arbitrarily reduce the ZZH coupling since we need this coupling to regulate $W_L W_L$ in an EW theory. Non-minimal scenarios can decouple certain Higgs states but there are so-called sum rules which guarantee that some states should be visible. Therefore a Higgs discovery cannot escape to ILC.

Figures 2 and 3 illustrate the robustness of ILC searches. In figure 2 one assumes a SM scenario for which a 120 GeV Higgs is observed, at 350 GeV centre of mass energy, in the HZ state with H dominantly decaying into b quarks. Note that the excellent ratio signal/background, s/b, allows a reduction in cross section by two orders of magnitude.

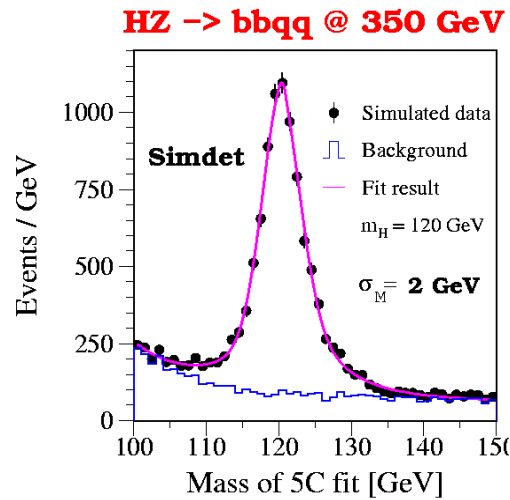

Figure 2: Higgs mass with jets at ILC.

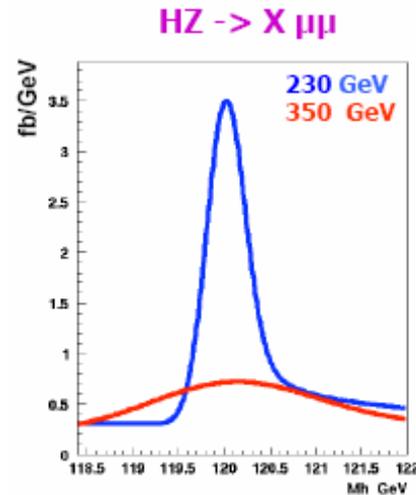

Figure 3: Higgs signal at ILC with di-muons at Ecm=230 and 350 GeV, radiation (ISR+FSR+BS) and ZZ background included.

In figure 3 one also starts from the HZ final state but this time makes no assumption on the decay mode of H but instead use the leptonic decay of Z to observe a Higgs signal by reconstructing the recoil mass to Z. Although with a reduced cross-section this method also leads to an excellent s/b, by operating at a centre of mass energy near threshold [2].

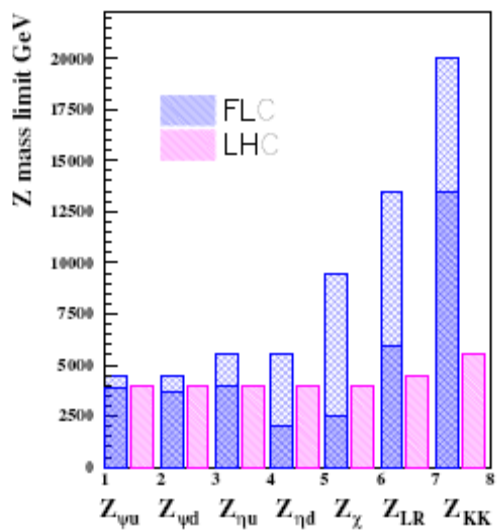

Figure 4: Z' mass limits at ILC (dark blue limits are from GigaZ with Z-Z' mixing).

If the absence of a Higgs signal is confirmed one can envisage two possible scenarios:

- with extra dimensions there is a family of Kaluza-Klein, KK, gauge bosons which replace the Higgs boson to cancel the $W_L W_L$ divergences
- in the absence of a Higgs boson $W_L W_L$ final states become strongly interacting (SI)

In the first scenario ILC sensitivity to Z'/KK particles [3] covers 5-20 TeV depending on the scenario as shown in figure 4. Below 5 TeV LHC provides the mass as an input and ILC allows to understand the origin by measuring V and A couplings precisely.

In the second scenario there should be deviations due to strong interactions in $W_L W_L$ final states. These deviations are in general observable on quartic couplings with WWνν or ZWW final states. This type of analysis requires W/Z separation which can be achieved with detectors [4] considered in ILC as shown in figure 5. It is also true that the quartic couplings are best measured at 1 TeV which gives the needed sensitivity [5] to insure visibility (figure 6). LHC can also observe these effects but this requires luminosities which won't be achieved at an early stage.

If there is a ρ type resonance then it can be observed in the reaction e+e- ->WW and already at 500 GeV can ILC provide a sufficient sensitivity to observe significant deviations [6] as shown on figure 7.

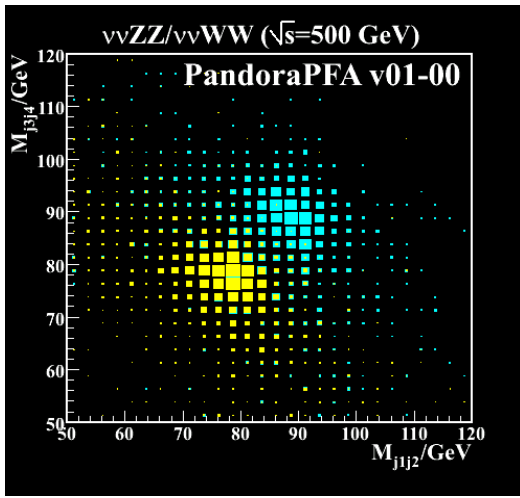

Figure 5: Hadronic mass separation for WWνν (yellow) and ZZνν (blue).

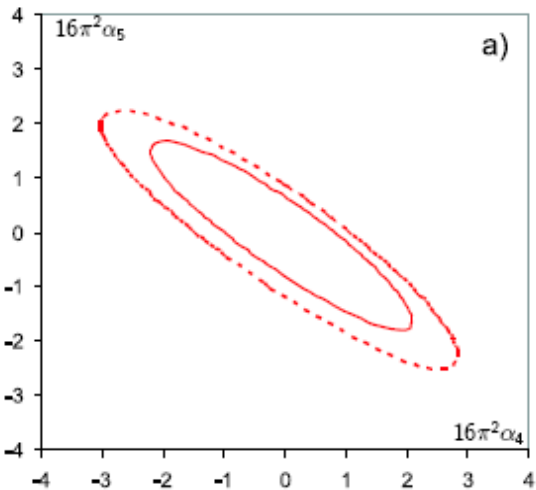

Figure 6: ILC limits (68% and 90% CL) on $\alpha_5$ and $\alpha_6$ with 1 ab$^{-1}$ at 1 TeV.

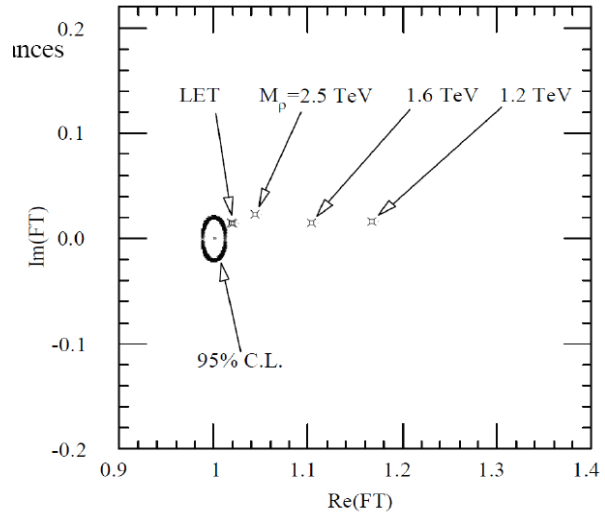

Figure 7: ILC sensitivity on ρ–like resonances with 500 fb$^{-1}$ at 500 GeV.

In conclusion scenario A although very difficult politically for ILC can well be defended scientifically. In the strongly interacting scenario ILC @1 TeV would, in some cases, be clearly superior but it will take quite some time to get the first significant answers from LHC. In the Higgsless scenarios with ρ–type resonances or KK recurrences elastically coupled to e+e-, ILC at 500 GeV goes beyond the mass sensitivity of LHC and with a polarized beam provides the tools to measure the vector and axial parts of these new couplings and therefore the origin of the effect.

One should finally recall that, precision measurements, PM, do not favour such a scenario but rather SM or MSSM.

## Scenario B

This scenario is sometimes called the 'theorist nightmare': Nature would provide a Higgs and nothing else up to the GUT scale. There are of course many reasons to think that this will not happen some of them purely theoretical (the mass hierarchy problem, the requirement of unification between strong and electroweak forces not achieved within the SM), others based on cosmological observations. How about PM from LEP/SLC and Tevatron? If the SM remains valid up to the GUT scale, theory predicts that 140 GeV<$m_H$<175 GeV which is not favoured by data as can be seen on figure 8 which combines the top mass measurements from Tevatron with the W mass measured at LEP+Tevatron. Without assuming the GUT prediction, LEP2 excludes at the 68% level the SM since from direct searches $m_H$>114.5 GeV, while the most probable value ~80 GeV. While not yet significant, this effect suggests that there are extra contributions which could, within MSSM, be provided by a moderately light stop component.

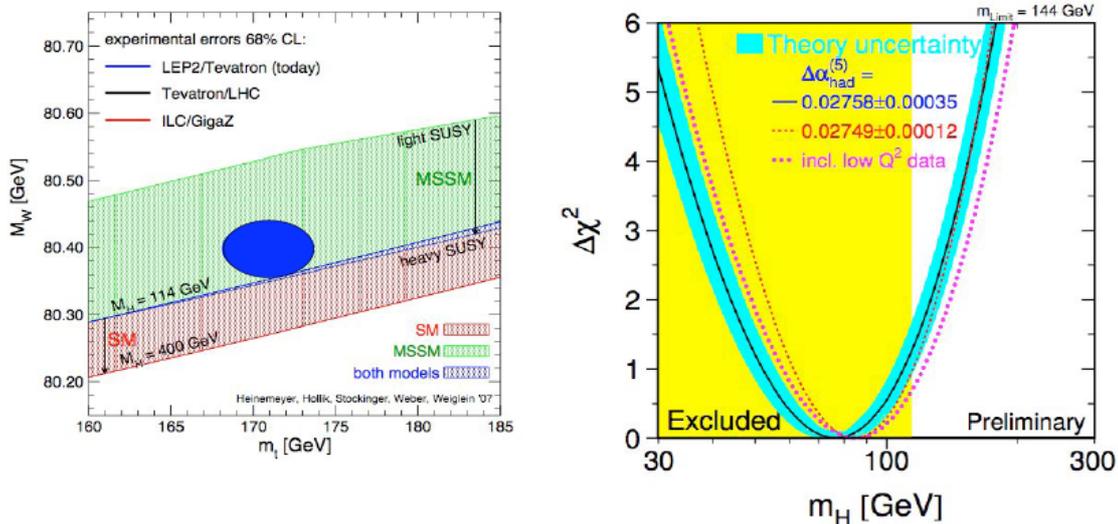

*Figure 8: Higgs mass predictions versus top and W mass (left). The green part is for MMSM and the red one for the SM. $\chi^2$ dependence of the overall EW fit right).*

The Higgs mass can also be predicted by measuring $\sin^2\theta_W$ and we will discuss in scenario C the resulting predictions.

LHC can discover such a SM Higgs particle with mass above 114 GeV. With limited accuracy however LHC may be unable to rule out the purely SM interpretation in the absence of new other signals. ILC has ten times more precision and a wider number of measurable channels, in particular ZHH and ttH very difficult if not impossible at LHC.

Figure 9 recalls this impressive set of measurements achievable [5] at ILC.
The GigaZ option, ILC running at the Z pole, would allow improving by more than one order of magnitude the accuracies reached at LEP1. It would then be possible to

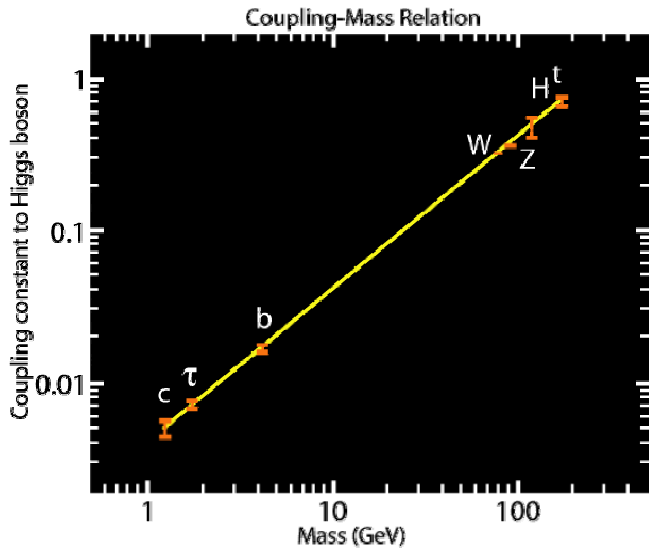

*Figure 9: Higgs couplings accuracies at ILC.*

narrow down the indirect prediction on the Higgs mass within the SM and therefore check, at the quantum level, the overall consistency of the SM. One should establish if:

**MHdirect=MHindirect ±5 GeV ?**

Further important tests of the Higgs SM are possible with ILC:
- Test of CPV (CP violation)
- Search for invisible decays at the % level

$\tau\tau$ decays provide the necessary observables to detect CPV violation. At LEP1 it was shown that polarisation of a $\tau$ can be efficiently measured from the hadronic decay modes. Here we need to correlate the polarisations of the two $\tau$ leptons which may cause certain problems given that the Higgs boson does not decay in its rest frame but preliminary studies indicate that these problems can be overcome [7].

Since a light Higgs boson, say below 150 GeV, couples very weakly to standard fermions it can easily receive a measurable branching ratio from any of the non standard extensions of the SM which predict light particles coupled to Higgs bosons. In several of these extensions one predicts significant, if not dominant, couplings to invisible particles (Majorons, sterile neutrinos etc…). It is therefore essential to provide the highest sensitivity on the measurement of the Higgs invisible branching ratio $BR_{inv}$, as it carries a large discovery potential.

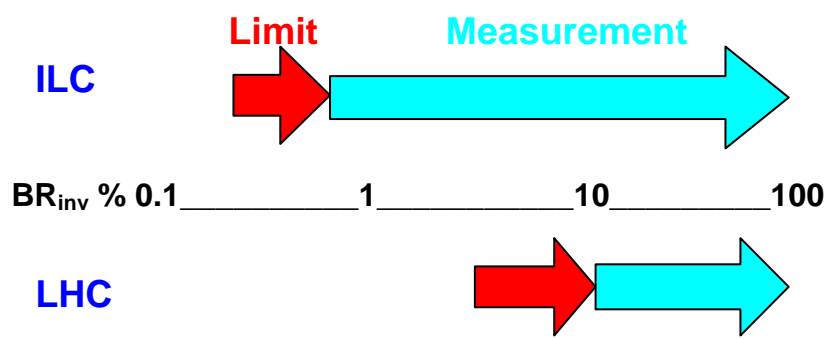

In the figure shown above [2] one sees that ILC extends very significantly the reach which could be achieved at LHC.

In conclusion LHC with limited accuracies on a limited set of measurements could be inconclusive for scenario B. Only ILC can ultimately tell if Higgs properties are consistent with a SM and 'nothing else'.

## Scenario C

In this scenario a heavy Higgs would be observed and nothing else. This Higgs boson would decay into ZZ and therefore be soon discovered at LHC. Furthermore, if mH>180 GeV, SUSY would seem excluded while one would need indirect contributions from new physics to explain PM from LEP/SLC and Tevatron.

In such a scenario ILC would play a very different role than for scenario B since fermionic branching ratios become negligible. The emphasis would therefore not be anymore on measuring Higgs decays but rather on measuring electroweak couplings Zff, ZWW and ZZH to detect indirectly the new physics at work.

To illustrate this scenario let us assume that the underlying model has extra dimensions and more specifically let us assume a Randall Sundrum, RS, scheme with so-called warped extra dimension. This model allows accommodating the hierarchy problem by assuming that there is exponential damping between a Planck brane and a TeV brane. It further allows explaining the mass hierarchy observed for fermions, from neutrino masses to the top mass, by assuming different localisations of the fermions in the $5^{th}$ dimension between these two branes. One assumes that the Higgs boson sits on the TeV brane, hence it's decoupling from the Planck scale. The top quark would be localized near the TeV brane while lighter fermions would come near the Planck brane.

These different localisations would have some observable consequences due to the KK bosons which would interact differently with the various fermions. In particular one could expect different electroweak couplings for the heaviest quarks.

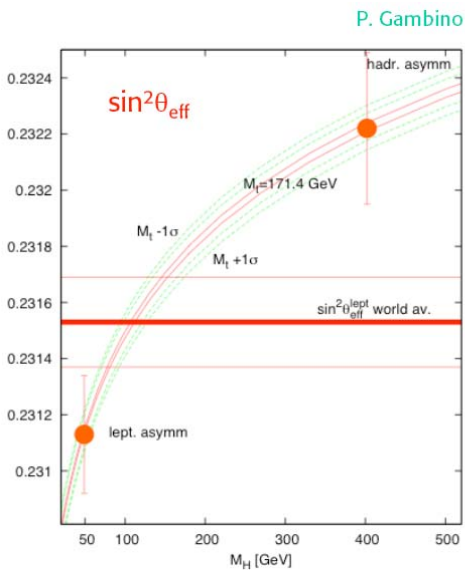

Figure 10: $sin^2\theta_{W\ dependence}$ with mH. The $1^{st}$ point comes from lepton the $2^{nd}$ from FB asymmetry with bb.

In this respect it is worth recalling the intriguing discrepancy observed in the most precise determinations of $sin^2\theta_W$ at LEP1 and SLC. For the latter one uses, with polarized electrons, the left-right asymmetry while LEP1 has the most precise determination through the forward backward asymmetry using b quarks and there is a ~3.5σ discrepancy between these two measurement.

The consequences of this discrepancy are shown in figure 10 where one can see that the leptonic asymmetry (dominated by the SLC result) predicts a very light Higgs (comparable to the W result of figure 8) which is inconsistent with the b measurement [8].

The final puzzle comes from the absence of deviation observed on Rb=Γb/Γhad.

One can reproduce [9] such features within the RS scheme assuming that there is a KK boson, Z', with mass ~3 TeV and by adjusting the respective 'positions' of the $b_R$ and $b_L$ quarks with respect to the TeV brane as shown in figure11.

Within this type of solution, very large deviations [9] are expected for top physics as shown in figure 12. Finally since the left right asymmetry can be measured at better than a %, ILC is precise enough to detect a Z' boson with a mass up to 20 TeV, therefore covering the whole 'reasonable' range of parameters for this model.

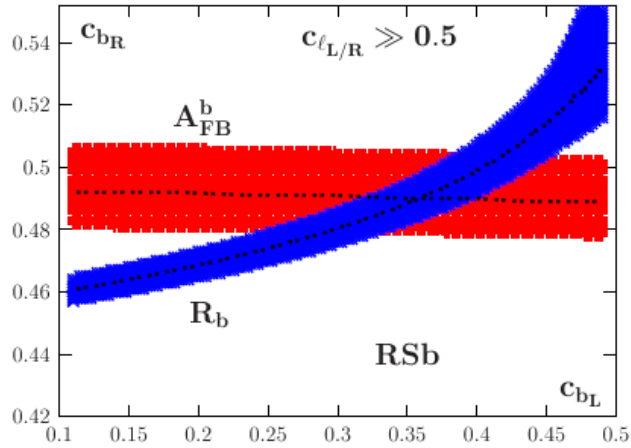 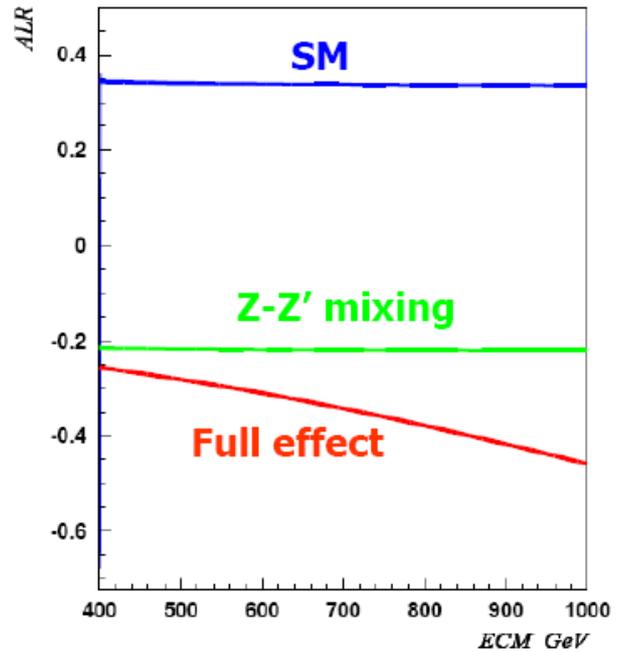

Figure 11: Contour plots giving $b_R$ and $b_L$ RS parameters consistent with Rb and the FB asymmetry.

Figure 12: $A_{LR}$ for top quarks for SM (blue) and RS (red). The green curve is due to Z-Z' mixing.

After reconciling the b asymmetry result on $\sin^2\theta_W$ with the leptonic results and Mw one still needs to explain the low Higgs mass prediction in apparent contradiction with the LEP2 limit. This in fact can be easily achieved since the RS model contains the needed ingredients to create the necessary inputs on the T and S variables to be consistent with a heavy Higgs [10].

This type of model also provides a EWSB mechanism where the Higgs boson appears as a Goldstone boson from a SI hidden sector. This is the so called strongly interacting light Higgs discussed in [11]. This scheme allows passing PM constraints but leaves significant imprints visible at LHC (through KK resonances) and/or ILC through deviations of the various Higgs couplings. LHC can directly discover such resonances up to 3 TeV while the indirect reach of ILC is ~8 TeV.

In conclusion above examples amply illustrate the high potential of ILC for scenario C.

## Scenario D

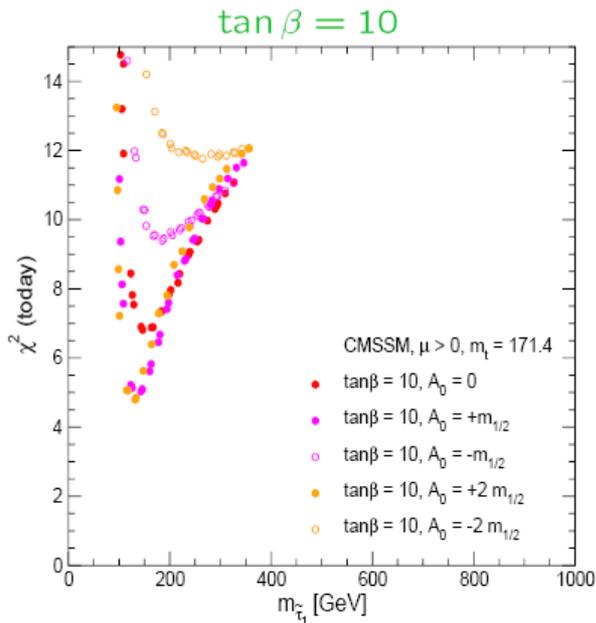

Figure 13: $\chi^2$ dependence of the EW fit with the stau mass for various SUSY sets.

In this scenario at least one Higgs boson would be observed with extra signals incompatible with SM interpretations. Our favourite choice at the present conference is SUSY and there are indeed several indications of light SUSY which are mainly coming from the W mass measurement combined with the top mass (see figure 8) and with the deviation observed on $(g-2)_\mu$ at the 3.5σ level. In [12] a fit was performed which predicts, in particular, light staus observable at ILC as shown in figure 13. Although not overwhelming these indications predict a wealth of exciting results which should come out quite soon from LHC giving further informations on the reach of ILC.

There are however a few caveats which need to be recalled. The limit from LEP2 MH>114.5 GeV excludes a large fraction of SUSY parameters provoking some concerns at the theoretical level about fine tuning. Recall however that within MSSM the true mass limit is Mh>90 GeV (and even much lower with CPV). There is even a slight indication [13] at LEP2 below 100 GeV as shown in figure 14. This indication would be consistent with MSSM if h/A/H have similar masses. A complex situation may occur if h/A/H are mass degenerate [14] and can mix with CPV as shown in figure 15. It will take ILC mass resolution and purity to disentangle this complicated scenario.

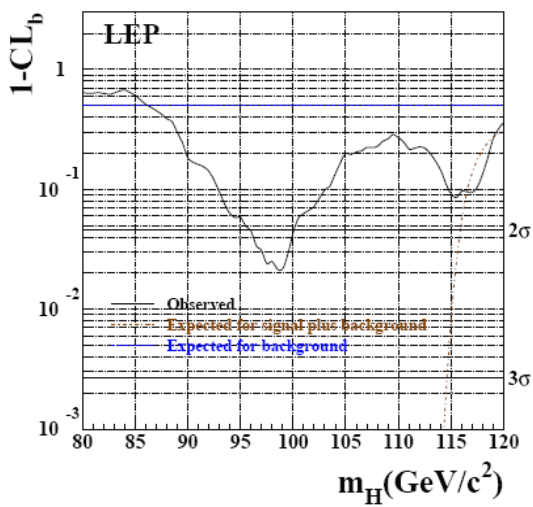

Figure 14: Background CL dependence versus mH observed at LEP2.

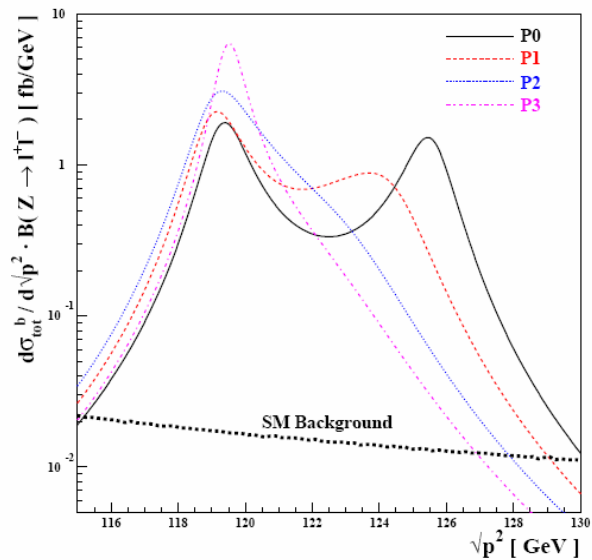

Figure 15: µµ recoil mass expected at ILC with a CPV scenario where h/H/A are quasi degenerate in mass.

Since ILC has excellent s/b for sleptons and gauginos it can provide, as well known, excellent and precise inputs to extract the fundamental SUSY parameters in conjunction with LHC. In particular while LHC can measure mass differences it has limited capabilities to determine absolute masses. ILC with polarization and threshold scans will offer dramatic improvements in the slepton and gaugino sectors in particular in determining the LSP mass. These features will allow reaching the accuracies needed to test the theory at the GUT scale.

This could have dramatic consequences in the neutrino sector [15] within SUSY with SO(10) as displayed in figure 16. From light slepton masses ILC could accurately predict the mass of the Majorana neutrino conveying the see-saw mechanism and also predict the absolute mass of the neutrino as displayed in this figure.

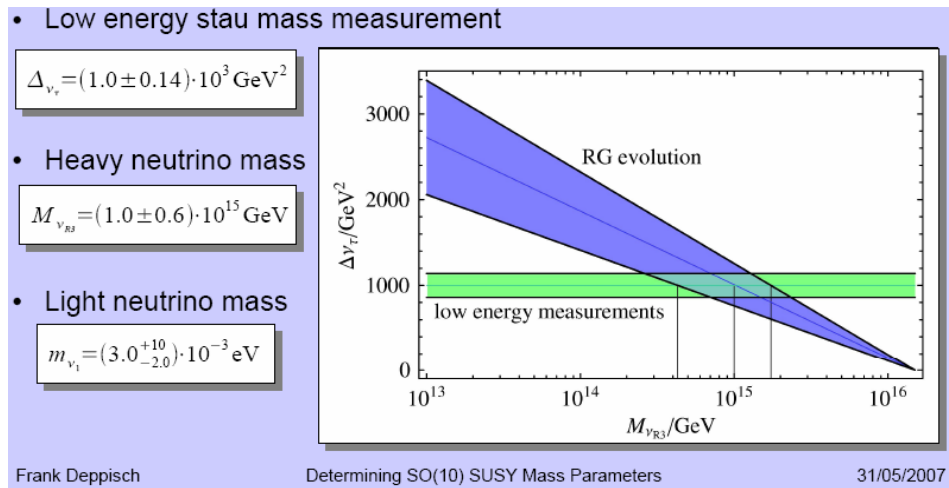

*Figure 16: Heavy (and light) neutrino mass determination using slepton accuracy measurements at ILC.*

As pointed out in [16] there are some blind regions in the SUSY mass spectrum which may compromise elucidation of the so-called LHC-1 problem. This occurs primarily in mass degenerate scenarios which may occur in certain DM as discussed below. Recall also that the meaning of 'mass degenerate' at LHC covers quite a large range. If one considers for instance a scenario, not unlikely, for which the lightest squark is a stop quark which would decay into $c\chi$, it would require a mass difference larger than 50 GeV between the stop mass and the neutralino mass to observe this signal.

At ILC the limitation comes from the $\gamma\gamma$ background and can be handled if the mass difference Dm>3 GeV as was shown for stau decaying into $\tau\chi$ for the co-annihilation DM scenario analysed in [17]. Needless to say that such analysis relies on efficient vetoing in the forward region of the detector which has received great attention.

To illustrate these features of ILC figure 17 shows the quality of the s/b separation for a DM solution given in [17] (so called point D where mstau=217 GeV Dm=5 GeV). This result comes from an update [18] shown at LCWS07 and demonstrates that an accuracy of ~0.1 GeV is achievable on Dm which is sufficient to predict the DM content of the universe at the WMAP/Planck accuracy level (see figure 18).

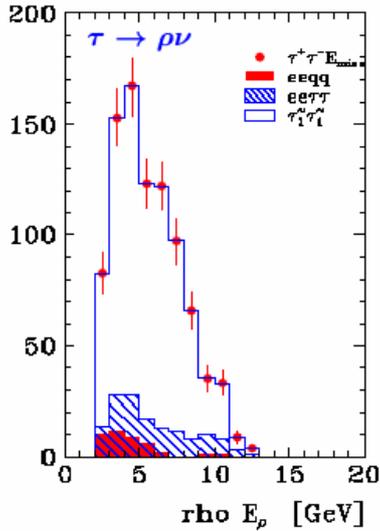 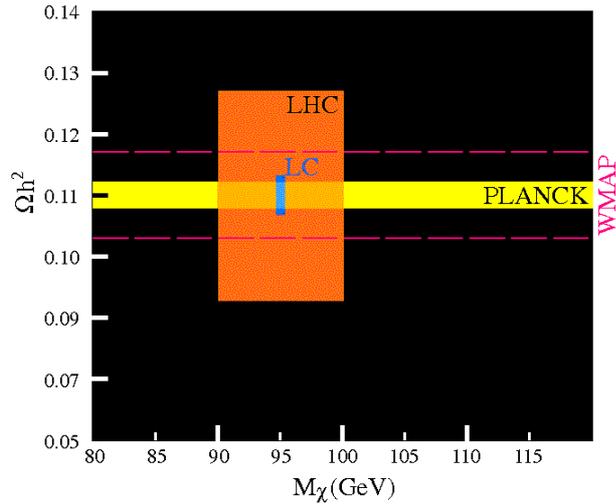

*Figure 17: $\rho$ energy reconstruction in $\tau\tau$ events with missing energy.*

*Figure 18: ILC predictions for DM in co-annihilation scenarios compared to satellite determinations.*

## Summary and conclusions

- ILC should, in some cases, complete LHC exploration of the **Terascale** and, in other cases, uniquely extend this exploration
- For the Higgs sector, SM or SUSY, ILC provides a superior reach for fundamental measurements and allows a full coverage of scenarios
- Measuring the top EW couplings at the ‰ offers full exploration in several extensions of the SM
- ILC together with LHC can fully reconstruct the underlying parameters of SUSY, allowing GUT extrapolations very promising in the leptonic sector
- ILC allows to cover SUSY 'mass degenerate' cases which are likely to occur in some DM scenarios

**Acknowledgements:** It is a pleasure to thank the organizers of SUSY2010 and in particular Sachio Komamiya for inviting me to this nice conference.